\documentclass[aps,twocolumn,amsmath,prx,showkeys]{revtex4-1}
\usepackage{graphicx}
\usepackage{subfigure}
\usepackage{amsmath}
\usepackage{bibentry}

\begin{document}

\title{Each state in a one-dimensional disordered system has two localization
lengths when the Hilbert space is constrained }

\author{Ye Xiong} 
\email{xiongye@njnu.edu.cn}
\affiliation{Institute of Theoretical Physics, Nanjing Normal University, Nanjing 210023,
P. R. China}

\begin{abstract} In disordered systems, the amplitudes of the localized states
will decrease exponentially away from their centers and the localization lengths
are characterizing such decreasing. In this article, we find a model in which
each eigenstate is decreasing at two distinct rates. The model is a
one-dimensional disordered system with a constrained Hilbert space: all
eigenstates $|\Psi\rangle$s should be orthogonal to a state $|\Phi \rangle$,
$\langle \Phi | \Psi \rangle =0$, where $|\Phi \rangle$ is a given exponentially
localized state. Although the dimension of the Hilbert space is only reduced by
$1$, the amplitude of each state will decrease at one rate near its center and
at another rate in the rest region, as shown in Fig.  \ref{fig1}.  Depending on
$| \Phi \rangle$, it is also possible that all states are changed from
localized states to extended states. In such a case, the level spacing
distribution is different from that of the three well-known ensembles of the
random matrices. This indicates that a new ensemble of random matrices exists
in this model. Finally we discuss the physics behind such phenomena and propose an
experiment to observe them. 

\end{abstract}

\maketitle

\section{Introductions} 

In Anderson localization, every localized state is characterized by one quantity
called the localization length $\lambda$. It describes how much the state is
localized in the real space, $|\Psi_E({\vec r})|\sim
e^{-|\vec{r}-\vec{r}_0|/\lambda(E)}$\cite{RevModPhys.80.1355}, where $r_0$ is
the center of the state at which the wavefunction takes the maximum value and
$E$ is the eigenenergy. This ansatz still applies thoroughly in the Anderson
disordered models in recent works, from one-dimension(1D) to three
dimension\cite{DeTomasi2016,Lin2015,Wang2015a,Belitz2016}, from the localized
states to the extended states with $\lambda=0$\cite{Yusipov2016,
PhysRevB.76.214204,Sheinfux2017,Pasek2016, Delande2017, DiSante2016a, Tikhonov2016a,
Garcia-Mata2016, Smith2017, Murphy2017,RevModPhys.67.357,Lee1981} and from Gaussian
orthogonal ensemble (GOE) to Gaussian symplectic ensemble (GSE) of random
matrices\cite{RevModPhys.69.731,RevModPhys.53.385}.

\begin{figure}[ht] 
  \includegraphics[width=0.45\textwidth]{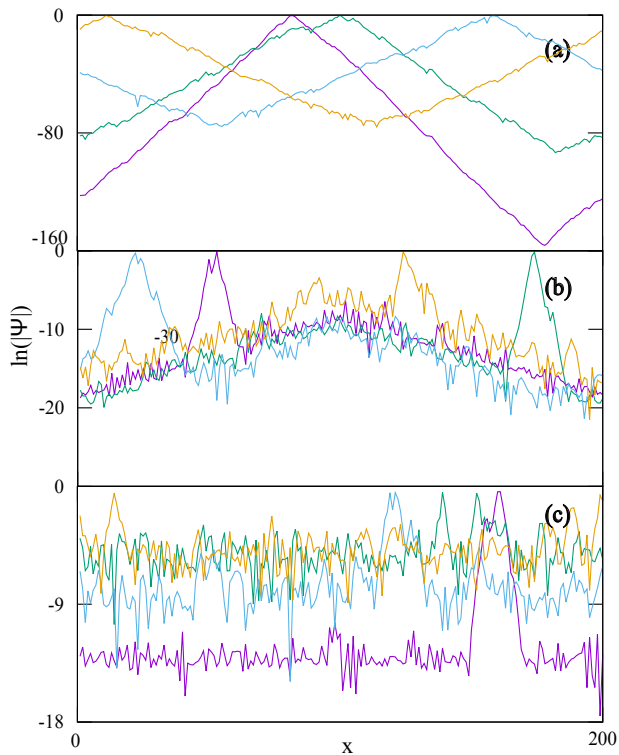}
  \caption[Fig]{\label{fig1} (a) The eigenstates in the real space for the
  Anderson  disordered model. (b) and (c) Those for the model with CHS.  For the
  sake of clarity, only $4$ typical states are plotted and the other
  states are similar. Each state in (b) has two localization lengths while it
  is an extended state in (c). The length of the ring is $N=200$, the strength of
  the disorder is $W=10$ and $\alpha$ is
  $0.1$ in (b) and $0$ in (c).}
\end{figure}

But in this article, we find that this ansatz may be broken down when an extra
constraint is subjected to the Hilbert space, $\langle \Phi | \Psi \rangle = 0$.
In another word, the dimension of the effective Hilbert space spanned by $|\Psi
\rangle$s is reduced by $1$ because $| \Phi \rangle$ is not in such Hilbert
space. Erasing a site at $i$ from a lattice is an example of such a constraint.
Here $|\Phi \rangle =|i\rangle$ and the effective Hilbert space is on the rest
lattice. In this article, the $|\Phi \rangle$ we are considering is an exponentially
decreasing function, which is neither the eigenstate of the Hamiltonian nor the
eigenstate of the position operator. We study the eigenstates of a 1D disordered
lattice in such constrained Hilbert space(CHS) and find that each state needs
two localization lengths to characterize its localization. As Fig.
\ref{fig1}(b) shown, the states first exponentially decrease faster near their
centers and then their decreasing change to a slower rate. When $\alpha=0$, as
shown in Fig.  \ref{fig1}(c), all states become extended, regardless of how
strong the disorder is in such 1D system.

The article is organized as follows: we first write down the static
Schr{\"o}dinger equation for a 1D disordered lattice in CHS. Interestingly, such
an equation is changed from homogeneous to non-homogeneous. The equation is
solved numerically on a finite lattice to show the eigenstates. Then the
transfer matrix method is developed to take part in the non-homogeneous term. We
argued that the ensemble of the orthogonal states during the QR decomposition
should be reexamined to pick up the correct Lyapunov exponents. The distribution
of the level spacing is also distinct from that of the traditional disordered
systems. Finally, we discuss how to realize such constraint in an experiment. 

\section{The non-homogeneous Schr{\"o}dinger equation in CHS}

We consider the Hamiltonian for a 1D Anderson disordered lattice with the on-site
disorders, 
\begin{equation}
H_{D} = \sum_{x=1}^{N} ( |x \rangle \langle
x+1| + h.c.) + \sum_{x=1}^{N} \epsilon_x |x \rangle \langle x|,
\end{equation}
where $\epsilon_x$ are the random numbers within $[-W/2,W/2]$, $x$ is the index
of the site on the 1D lattice and the integer $N$ is the length. We take the
periodic boundary condition in the calculations. 

It is well-known that the standard static Schr{\"o}dinger equation $E|\Psi
\rangle = H_{D}
| \Psi \rangle$ can be obtained by minimizing $\langle \Psi |H_D| \Psi \rangle$
under the constraint $\langle \Psi| \Psi \rangle=1$. The eigenenergy $E$ is a
Lagrange multiplier. Now as we have one more constraint $\langle \Psi | \Phi
\rangle=0$, the static Schr{\"o}dinger equation changes to
\begin{eqnarray}
E | \Psi \rangle &= & H_{D} | \Psi \rangle + \mu | \Phi \rangle ,\label{Hc} \\
\langle \Phi | \Psi \rangle &=&0.
\end{eqnarray}
Here $\mu$ is another Lagrange multiplier and it should take the value to make
$\langle \Phi | \Psi \rangle =0$. In the section on the experimental setup, we
will give another argument to prove the validity of the above equations.

The eigenvalues $E$ and the corresponding eigenfunctions $|\Psi \rangle$ can be
found by solving a general eigenproblem
\begin{equation}
\begin{pmatrix} EI & 0 \\ 0 & 0 \end{pmatrix} \begin{pmatrix} |\Psi \rangle \\
\mu \end{pmatrix} = \begin{pmatrix} H_D & |\Phi \rangle \\ \langle \Phi | & 0
\end{pmatrix} \begin{pmatrix} |\Psi \rangle \\ \mu \end{pmatrix},
\label{Hn}
\end{equation}
where $I$ is a $N\times N$ identity matrix. It is also equivalent to an
eigenproblem for a defective non-hermitian matrix\cite{2021arXiv210400847X}.

We plot several eigenstates $\Psi(x)=\langle x | \Psi \rangle$ in Fig.
\ref{fig1}(b). The strength of disorder is $W=10$ and the wavefunction in the
constraint is $\Phi(x) = e^{-\alpha|x-x_0|}$ whose center $x_0$ is at the center
of the ring. We use the GEM package in {\it octave} to perform the calculation
with much higher precision so that the results have not been smeared out by the
round-off errors.  For the sake of comparative analysis, we first plot the
traditional eigenstates of $H_D$ in Fig. \ref{fig1}(a). Every state is
decreasing exponentially over the whole chain at a constant rate, $\frac{1}{\xi(E)}$.  In
Fig. \ref{fig1}(b), the wavefunctions first decrease rapidly around
their centers and then change to decrease/increase at a slower rate. After
fitting the data, we find this slower rate is $\alpha$, which is independent of
the eigenenergies $E$ and the central positions of the wavefunctions.
When $\alpha=0$, as shown in Fig. \ref{fig1}(c), all states become extended,
regardless of the strength of the disorder. 

In traditional 1D disordered models, as the states $\Psi_0(x)$s are
exponentially localized, the local environments at a far distance should not
affect the states.  The overlaps with $\Phi(x)$, $\langle \Phi | \Psi_0 \rangle
\sim \max\{ e^{-\alpha d}, e^{-\frac{1}{\lambda} d}\}$, are exponentially small
but are not exactly zero. Here $d$ is the distance between the centers of
$\Psi_0(x)$ and $\Phi(x)$. To make such overlap be exactly zero, as the
constraint requires, the state $\Psi(x)$ must be disturbed from $\Psi_0(x)$ in
the scale of $\max\{e^{-\alpha d}, e^{-\frac{1}{\lambda} d}\}$. This is
confirmed in our calculations as the Lagrange multiplier $\mu \sim
\max\{e^{-\alpha d}, e^{-\frac{1}{\lambda} d}\}$. So an exponentially small
factor $\mu$ in the non-homogeneous Schr{\"o}dinger equation Eq. \ref{Hc} is
important and can affect the character of localization in the far distance.
This is distinct from many equations in which an exponentially small term is
ignorable. To confirm this conclusion, we employ the transfer matrix method to
study the localization lengths of such a system.

\section{The transfer matrix method in CHS}

The traditional transfer matrix method is used to calculate the Lyapunov
exponents of the transfer matrix that is relating the wavefunctions on the $L$th
and the $(L+1)$th slices with those on the $0$th and the $1$st
slices\cite{PhysRevLett.47.1546,Pichard_1981,MacKinnon1983} . It is
subjected to the homogeneous Schr{\"o}dinger equation and only the diagonal
elements of the $R$ matrix during the $QR$ decomposition are
interested\cite{PhysRevB.104.104203}. Here we
first extend the method to the non-homogeneous case and argue that the
orthogonalized vectors in the $Q$ matrix should be inspected first.

\begin{figure}[ht] 
  \includegraphics[width=0.45\textwidth]{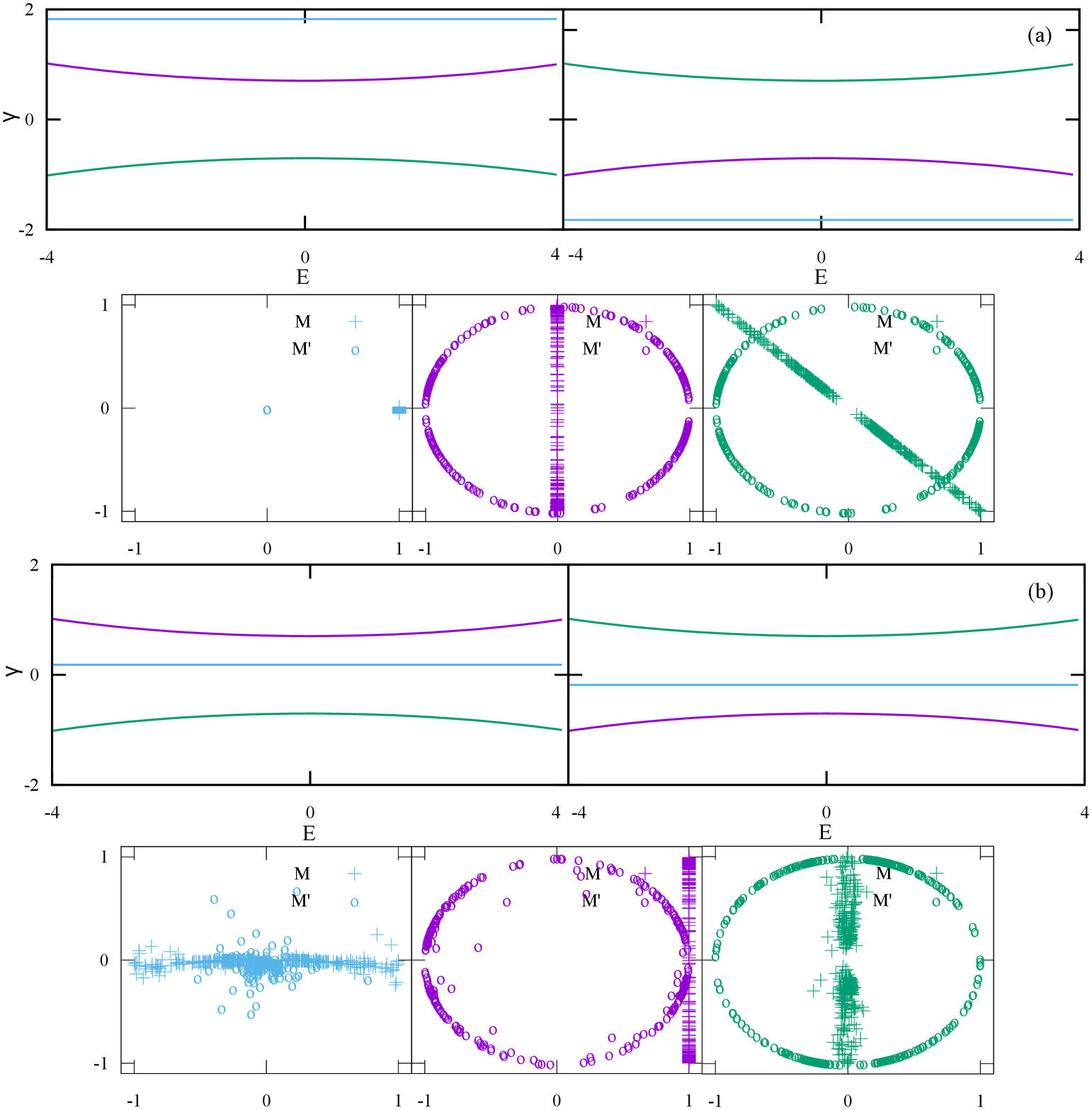}

  \caption[Fig]{\label{fig2} (a) The three Lyapunov exponents, $\alpha$ and $\pm
  \frac{1}{\lambda(E)}$ ($-\alpha$ and $\mp \frac{1}{\lambda(E)}$) as the
  function of energy $E$ for the transfer matrix $M$ (left) and $M'$ (right).
  For each Lyapunov exponent $\gamma$ on the left, there is its partner
  $-\gamma$ on the right. The first two components of 
  the orthonormal columns in $Q$, $\vec{z}_i$s, are plotted below for the
  three pairs of Lyapunov exponents. There are $1000$ disorder configurations being
  considered and their $\vec{z}_i$s are all plotted in the figures. 
  The colors of the points are used to distinguish the corresponding Lyapunov exponent
  and the ``plus'' and the ``circle'' symbols are for $M$ and $M'$,
  respectively. Here $\alpha=1.83 > \frac{1}{\lambda(E)}$, $W=10$ and $E=1$.
  As there is no overlap between $\{\vec{z}_i\}$s for the pair
  $(\alpha,-\alpha)$, $\alpha$ is not the true exponent to describe the
  asymptotic behaviors of the states. (b) When $\alpha=0.18 <
  \frac{1}{\lambda(E)}$, the overlaps emerge so that both $\alpha$ and
  $\frac{1}{\lambda(E)}$ are describing the asymptotic behaviors of the states
  simultaneously.}

\end{figure} 

We redefine the transfer matrix $M_x(E)$ as the square matrix in  
\begin{equation}
	\begin{pmatrix} \Psi_{x+1} \\ \Psi_{x} \\ \Phi_{x+R+1} \end{pmatrix}
	=\begin{pmatrix}
		E-\epsilon_x & -1 & \mu e^{\alpha R} \\
		1 & 0 & 0 \\
		0 & 0 & e^{-\alpha} 
	\end{pmatrix}
	\begin{pmatrix} \Psi_{x} \\ \Psi_{x-1} \\ \Phi_{x+R} \end{pmatrix},
	\label{Tr}
\end{equation}
where $\Psi_{x}$ is the wavefunctions on  the $x$th slice (the $x$th lattice in
our 1D model), $E$ is the energy, $\Phi_{x+R}$ is the wavefunction of $|\Phi
\rangle$ on the $(x+R)$th site. Here $\Phi$ is offset $R$ sites so that $\mu
e^{\alpha R}$ is in the scale of the unit even when $\mu$ is exponentially small.
As $\mu$ has been determined by the constraint but the exact value of $\mu$
is not important in this treatment, the constraint is not needed to be written
down explicitly in the transfer matrix. In another word, after $\mu$ is
determined by the equation of the constraint, we can always choose a proper
offset $R$ to scale the term $\mu e^{\alpha R}$.

The transfer matrix $M=\prod_{x=1}^{L} M_x$ can be $QR$ decomposed into the
product of a matrix $Q$ having orthonormal columns and an upper triangular
matrix $R$. The diagonal elements of $R$ matrix are $e^{L\gamma_i}$, where
$\gamma_i$ are the Lyapunov exponents and the orthonormal columns in $Q$,
$\vec{z}_i$, are the corresponding typical states. Due to the variations of
the disordered configurations, each $\vec{z}_i$ actually forms a group of ensemble $\{
\vec{z}_i \}$. If the typical state $\vec{z}_i$ does exist for the
disordered chain, one must find consistent results regardless of viewing the chain
from the left to the right or from the right to the left. So the ensemble of $\{
\vec{z}_i \}$ for the Lyapunov exponent $\gamma_i$ must match or at least 
overlaps with the ensemble of $\{ \vec{z}'_i \}$ for the Lyapunov exponent
$-\gamma_i$ of the transfer matrix $M'=\prod_{x=L}^{1} M'_x$. Here the transfer
matrix $M'$ is counting from $x=L$(the right) to $x=1$ (the left) and $M'_x$ is
the same as $M_x$ except that the last element is replaced by $e^{\alpha}$
because the decreasing $\Phi(x)$ changes to the increasing function as reversing
$x$.

The three Lyapunov exponents of the transfer matrix $M$ are $\alpha$ and $\pm
\gamma(E)$, where $\pm \gamma(E)$ are the exponents of the disordered chain in
the absence of constraint. So it seems that the redefinition of the transfer
matrix only enrolls an additional exponent $\alpha$ which can be attributed to
the asymptotic behavior of $\Phi(x)$. But as shown in Fig. \ref{fig2}(a), some
of the exponents are untrue. We first plot the exponents as the function of the
energy $E$ for the transfer matrix $M$ and the transfer matrix $M'$,
respectively. For each exponent $\gamma_i$ for $M$, there is a $-\gamma_i$
exponent for $M'$. Then the ensemble of the corresponding typical states
$\vec{z}_i$ for such pairs of exponents are plotted. As these typical states are
normalized, only the first two components of the states are plotted. When
$\alpha>\frac{1}{\lambda(E)}$, it is shown that the ensembles of $\vec{z}$ and
$\vec{z}'$ for the pair ($\alpha$, $-\alpha$) do not overlap with each other. So
$\frac{1}{\alpha}$ is not the localization length of the model, because if the
corresponding typical state in $Q$ matrix does exist, it should be observed
independent of viewing from the left or the right. In such case, each state
is only characterized by one localization length, $\lambda(E)$. This conclusion
is reasonable because in the extreme case $\alpha \to \infty$, $\Phi(x)$ is a
delta function so the constraint is equivalent to the elimination of the
center site from the lattice.  This will only change the ring with $L$ sites to
a chain with $L-1$ sites. All bulk localized states are not affected by such
boundary condition so the localization behavior of the model is the same as
that of a traditional disordered model. But when $\alpha < \gamma(E)$, in Fig.
\ref{fig2}(b), the ensembles overlap so that both $\frac{1}{\alpha}$ and
$\lambda(E)$ are the localization lengths simultaneously. 

\section{The distribution of the level spacing}

\begin{figure}[ht] 
  \includegraphics[width=0.45\textwidth]{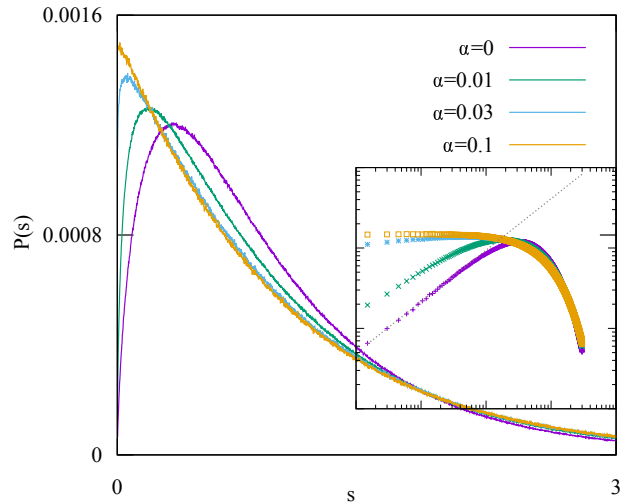}
  \caption[Fig]{\label{fig3} The distributions of the normalized level spacing $s$
  for $\alpha=0,0.01,0.03,0.1$, respectively. The inset shows the same
  distributions in
  logarithmic axes.  The straight line, $f(s) \sim s^{0.64}$, is a guide to
  the eye. The length of the ring is $N=600$, the strength of disorder is $W=10$ and $10^7$ samples are
  considered in the statistics.}
\end{figure} 

In the 1D Anderson disordered lattice, the distribution of the level spacing is
a Poisson curve $\sim e^{-s}$, where $s$ is the normalized nearest neighboring
eigenenergy spacing. The maximal distribution appears at $s=0$, which indicates that all
eigenstates are localized and cannot repulse the other states in the energy.
When the states become extended, such repulsion exercises so that the
distribution is changed to the Wigner distribution\cite{wigner_1951,Dyson,
Dyson1} 
\begin{equation}
P(s) \sim s^{\beta} e^{-\frac{\pi\beta^4}{4}s^2},
\label{wigner}
\end{equation}
where $\beta$ is $1$, $2$ and $4$ for the orthogonal, the unitary and the
symplectic ensembles, respectively. 

In Fig. \ref{fig3}, we show such distributions $P(s)$ when the constraint is
subjected. When $\alpha=0.1$, although the eigenstates are still characterized
by two localization lengths, the distribution is still a Poisson function. As
$\alpha$ is decreasing, one of the localization lengths becomes longer and
longer and dominates the asymptotic behaviors of the states.  So they become
more and more extended. This is confirmed by the fact that the distribution is
distinct from the Poisson function and becomes more and more Wigner-like. When
$\alpha=0$, all states should become extended and the distribution becomes a
Wigner function exactly. But interestingly, as the inset shows, the $\beta$ of
such Wigner function is $0.64$, a value distinct from those of the three
well-known ensembles. This indicates that the mechanics to delocalize the states
in this model are different from that of the competition between the disordered
potential and the kinetic energy.  We have no more discussions on the new ensemble
at this stage. But we think the emergence of the new ensemble is related to the
non-hermitian random matrix at an exceptional point because Eq. \ref{Hn} is
equivalent to the eigenproblem of a non-hermitian defective matrix.

\section{The experiments to observe the phenomena}

\begin{figure}[ht] 
  \includegraphics[width=0.45\textwidth]{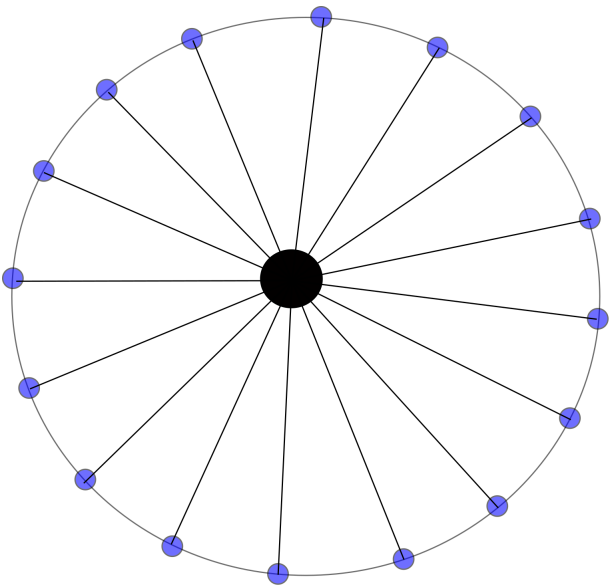}
  \caption[Fig]{\label{fig4} The electron on the quantum dots (blue points) can
  jump to the nearest-neighboring dots or the 
  central quantum dot (black point). The thin lines indicate
  the hoppings between the dots.}
\end{figure} 

The CHS can be realized in microscopic electronic systems, macroscopic
mechanical systems, or optical systems. We consider a ring of quantum dots
(resonant cavities) with effective hoppings between the nearest
neighboring dots (cavities). From now on, we will only focus on the system of quantum
dots but the mechanics are the same in the other systems. The disorder is introduced
by the on-site energies of the dots. There is another dot at the center of
the ring. Such a dot is weakly connected with all dots on the ring by an
effective hopping $t'$. The energy level of the center dot is detuned to be
$\delta$ away from the energy $E$ we are considering. After projecting out the
level of the center dot, the Hamiltonian for the dots on the ring is 
\begin{equation}
H_{\text{LR}} = H_{D} + \sum_{x,x'} \frac{t'^2}{\delta} |x \rangle \langle x'| ,
\label{Hlr}
\end{equation}
where $H_{D}$ is the disordered Hamiltonian and the
later term describes the effective hoppings between any pair of quantum dots 
mediated by the center dot. Such long-range hopping
term\cite{PhysRevLett.116.250402, PhysRevB.94.144206} can also be written as $
\frac{t'^2}{\delta}N |\Phi \rangle \langle \Phi |$, where $N$ is the number of
dots on the ring and $|\Phi \rangle = \frac{1}{\sqrt{N}} \sum_{x} | x \rangle$.
When $\frac{t'^2}{\delta}N$ is a huge number as compared to the energy scale of
$H_{D}$, the spectrum of $H_{\text{LR}}$ are composed by a band with $N-1$
levels and one level $|\Phi\rangle$ at the huge energy. Because the eigenstates of a hermitian
Hamiltonian are orthogonal to each other, the states in the band, $|\Psi
\rangle$s, must be orthogonal to the state $|\Phi\rangle$.
And such huge
gap between the single level and the band ensures us to ignore the single level
when the energy we are interested in is
in the scale of the band energy. Now the constraint $\langle \Psi | \Phi\rangle
=0$ must be subjected because the Hilbert space of the band states does not have
the state $|\Phi \rangle$. 

The above argument can be written down as
\begin{equation}
H_{\text{LR}} \sim (1-P) H_{D} (1-P) + \frac{t'^2}{\delta} N P |\Phi \rangle
\langle \Phi | P,
\label{pr}
\end{equation}
where the first-order perturbation theory pronounces that the off-diagonal
hoppings between the states in the band and the single level can be ignored and $P=
|\Phi \rangle \langle \Phi|$ is the project operator on the single level.
Such perturbation is approaching to be exact for huge gap. So the static
Schr\"{o}dinger equation $H_{\text{LR}}|\Psi\rangle =E|\Psi \rangle$ becomes   
\begin{equation}
H_D|\Psi \rangle - \langle \Phi |H_{D}| \Psi \rangle |\Phi \rangle =E |\Phi
\rangle,
\end{equation}
which is non-homogeneous. This equation is the same as Eq. \ref{Hc} which is
derived from the variational functional method. 

One can observe the transport behaviors on the ring. As all states are extended,
one may find a finite conductance in such a 1D strong disordered lattice. 

\section{Conclusions and discussions}

We have studied the Anderson localization for a 1D disordered lattice in CHS.
The constraint is $\langle \Psi |\Phi \rangle=0$ and $|\Phi \rangle$ is a state
that is decreasing in the space with the rate $\alpha$. When $\frac{1}{\alpha} >
\lambda(E)$, where $\lambda(E)$ is the native localization length of the state,
the state will need two localization lengths to characterize its shape in the
space. It will first exponentially decrease with a faster rate,
$\frac{1}{\lambda(E)}$, near its center and then changes to a slower rate, $\alpha$.
When $\alpha=0$, in every state, the slower rate dominates. So they become
extended in such a disordered system. 

Here we present another argument to prove the above conclusions. The solution
$\Psi(x)=\langle x |\Psi\rangle$ of the non-homogeneous differential equation,
Eq. \ref{Hc}, is
\begin{equation} \Psi(x)=\Psi_0(x) + \int dx' G(x,x')\Phi(x'), \end{equation}
where $G(x,x')$ is the green function $(E-H_D)G(x,x')=\delta(x-x')$ and
$\Psi_0(x)$ is the general solution, $(E-H_D) \Psi_0(x)=0$. The green function
is $G(x,x') \sim e^{-\frac{1}{\lambda(E)} |x-x'|}$ after averaged over the disorder
configurations. After Fourier transformation, the convolution becomes the product of
$G(k)$ and $\Phi(k)$. When $\alpha=0$, $\Phi(k)$ is a delta function at $k=0$
and so as for the product of $G(k)$ and $\Phi(k)$. As a result, $\Psi(x)$
becomes an extended state. 

In this article, we only consider the case $\Phi(x)=e^{-\alpha|x-x_0|}$. It will
be interesting to consider a staggered function, a random function, or a power-law 
decreasing function. One may find new localizations such as power-law 
localized states, real-space distinguished topological bands, or disordered
ensembles with new $\beta$s in these systems.

The mechanics of the model may be applied to many-body
localization\cite{RevModPhys.91.021001}. There are a series of conserved local
quantities in such systems. They can be considered as the native constraints. So
the rates at which these conserved quantities are localized should provide an
upper bound on the localization lengths. 

The systems in CHS are also junctions. They will connect the homogeneous
Schr\"{o}dinger equation for a long-range hopping Hamiltonian $H_{\text{LR}}$
with a non-homogeneous Schr\"{o}dinger equation for a short-range hopping
Hamiltonian $H_{D}$. They will connect the eigenproblem for a hermitian matrix
with that for a non-hermitian matrix at the exceptional point. A new route to
understand these problems may be found through this junction.

{Acknowledgments.---} 
The work was supported by the 
National Foundation of Natural Science in China Grant Nos. 10704040.

\bibliographystyle{apsrev4-1}
\bibliography{main}

\begin{thebibliography}{30}%
\makeatletter
\providecommand \@ifxundefined [1]{%
 \@ifx{#1\undefined}
}%
\providecommand \@ifnum [1]{%
 \ifnum #1\expandafter \@firstoftwo
 \else \expandafter \@secondoftwo
 \fi
}%
\providecommand \@ifx [1]{%
 \ifx #1\expandafter \@firstoftwo
 \else \expandafter \@secondoftwo
 \fi
}%
\providecommand \natexlab [1]{#1}%
\providecommand \enquote  [1]{``#1''}%
\providecommand \bibnamefont  [1]{#1}%
\providecommand \bibfnamefont [1]{#1}%
\providecommand \citenamefont [1]{#1}%
\providecommand \href@noop [0]{\@secondoftwo}%
\providecommand \href [0]{\begingroup \@sanitize@url \@href}%
\providecommand \@href[1]{\@@startlink{#1}\@@href}%
\providecommand \@@href[1]{\endgroup#1\@@endlink}%
\providecommand \@sanitize@url [0]{\catcode `\\12\catcode `\$12\catcode
  `\&12\catcode `\#12\catcode `\^12\catcode `\_12\catcode `\%12\relax}%
\providecommand \@@startlink[1]{}%
\providecommand \@@endlink[0]{}%
\providecommand \url  [0]{\begingroup\@sanitize@url \@url }%
\providecommand \@url [1]{\endgroup\@href {#1}{\urlprefix }}%
\providecommand \urlprefix  [0]{URL }%
\providecommand \Eprint [0]{\href }%
\providecommand \doibase [0]{http://dx.doi.org/}%
\providecommand \selectlanguage [0]{\@gobble}%
\providecommand \bibinfo  [0]{\@secondoftwo}%
\providecommand \bibfield  [0]{\@secondoftwo}%
\providecommand \translation [1]{[#1]}%
\providecommand \BibitemOpen [0]{}%
\providecommand \bibitemStop [0]{}%
\providecommand \bibitemNoStop [0]{.\EOS\space}%
\providecommand \EOS [0]{\spacefactor3000\relax}%
\providecommand \BibitemShut  [1]{\csname bibitem#1\endcsname}%
\let\auto@bib@innerbib\@empty
\bibitem [{\citenamefont {Evers}\ and\ \citenamefont
  {Mirlin}(2008)}]{RevModPhys.80.1355}%
  \BibitemOpen
  \bibfield  {author} {\bibinfo {author} {\bibfnamefont {F.}~\bibnamefont
  {Evers}}\ and\ \bibinfo {author} {\bibfnamefont {A.~D.}\ \bibnamefont
  {Mirlin}},\ }\href {\doibase 10.1103/RevModPhys.80.1355} {\bibfield
  {journal} {\bibinfo  {journal} {Rev. Mod. Phys.}\ }\textbf {\bibinfo {volume}
  {80}},\ \bibinfo {pages} {1355} (\bibinfo {year} {2008})}\BibitemShut
  {NoStop}%
\bibitem [{\citenamefont {{De Tomasi}}\ \emph {et~al.}(2016)\citenamefont {{De
  Tomasi}}, \citenamefont {Roy},\ and\ \citenamefont {Bera}}]{DeTomasi2016}%
  \BibitemOpen
  \bibfield  {author} {\bibinfo {author} {\bibfnamefont {G.}~\bibnamefont {{De
  Tomasi}}}, \bibinfo {author} {\bibfnamefont {S.}~\bibnamefont {Roy}}, \ and\
  \bibinfo {author} {\bibfnamefont {S.}~\bibnamefont {Bera}},\ }\href {\doibase
  10.1103/PhysRevB.94.144202} {\bibfield  {journal} {\bibinfo  {journal}
  {Physical Review B}\ }\textbf {\bibinfo {volume} {94}},\ \bibinfo {pages}
  {144202} (\bibinfo {year} {2016})}\BibitemShut {NoStop}%
\bibitem [{\citenamefont {Lin}\ and\ \citenamefont
  {Popovi{\'{c}}}(2015)}]{Lin2015}%
  \BibitemOpen
  \bibfield  {author} {\bibinfo {author} {\bibfnamefont {P.~V.}\ \bibnamefont
  {Lin}}\ and\ \bibinfo {author} {\bibfnamefont {D.}~\bibnamefont
  {Popovi{\'{c}}}},\ }\href {\doibase 10.1103/PhysRevLett.114.166401}
  {\bibfield  {journal} {\bibinfo  {journal} {Physical Review Letters}\
  }\textbf {\bibinfo {volume} {114}},\ \bibinfo {pages} {166401} (\bibinfo
  {year} {2015})}\BibitemShut {NoStop}%
\bibitem [{\citenamefont {Wang}\ \emph {et~al.}(2015)\citenamefont {Wang},
  \citenamefont {Su}, \citenamefont {Avishai}, \citenamefont {Meir},\ and\
  \citenamefont {Wang}}]{Wang2015a}%
  \BibitemOpen
  \bibfield  {author} {\bibinfo {author} {\bibfnamefont {C.}~\bibnamefont
  {Wang}}, \bibinfo {author} {\bibfnamefont {Y.}~\bibnamefont {Su}}, \bibinfo
  {author} {\bibfnamefont {Y.}~\bibnamefont {Avishai}}, \bibinfo {author}
  {\bibfnamefont {Y.}~\bibnamefont {Meir}}, \ and\ \bibinfo {author}
  {\bibfnamefont {X.~R.}\ \bibnamefont {Wang}},\ }\href {\doibase
  10.1103/PhysRevLett.114.096803} {\bibfield  {journal} {\bibinfo  {journal}
  {Physical Review Letters}\ }\textbf {\bibinfo {volume} {114}},\ \bibinfo
  {pages} {096803} (\bibinfo {year} {2015})}\BibitemShut {NoStop}%
\bibitem [{\citenamefont {Belitz}\ and\ \citenamefont
  {Kirkpatrick}(2016)}]{Belitz2016}%
  \BibitemOpen
  \bibfield  {author} {\bibinfo {author} {\bibfnamefont {D.}~\bibnamefont
  {Belitz}}\ and\ \bibinfo {author} {\bibfnamefont {T.~R.}\ \bibnamefont
  {Kirkpatrick}},\ }\href {\doibase 10.1103/PhysRevLett.117.236803} {\bibfield
  {journal} {\bibinfo  {journal} {Physical Review Letters}\ }\textbf {\bibinfo
  {volume} {117}},\ \bibinfo {pages} {236803} (\bibinfo {year}
  {2016})}\BibitemShut {NoStop}%
\bibitem [{\citenamefont {Yusipov}\ \emph {et~al.}(2017)\citenamefont
  {Yusipov}, \citenamefont {Laptyeva}, \citenamefont {Denisov},\ and\
  \citenamefont {Ivanchenko}}]{Yusipov2016}%
  \BibitemOpen
  \bibfield  {author} {\bibinfo {author} {\bibfnamefont {I.}~\bibnamefont
  {Yusipov}}, \bibinfo {author} {\bibfnamefont {T.}~\bibnamefont {Laptyeva}},
  \bibinfo {author} {\bibfnamefont {S.}~\bibnamefont {Denisov}}, \ and\
  \bibinfo {author} {\bibfnamefont {M.}~\bibnamefont {Ivanchenko}},\ }\href
  {\doibase 10.1103/PhysRevLett.118.070402} {\bibfield  {journal} {\bibinfo
  {journal} {Physical Review Letters}\ }\textbf {\bibinfo {volume} {118}},\
  \bibinfo {pages} {070402} (\bibinfo {year} {2017})}\BibitemShut {NoStop}%
\bibitem [{\citenamefont {Xiong}\ and\ \citenamefont
  {Xiong}(2007)}]{PhysRevB.76.214204}%
  \BibitemOpen
  \bibfield  {author} {\bibinfo {author} {\bibfnamefont {S.-J.}\ \bibnamefont
  {Xiong}}\ and\ \bibinfo {author} {\bibfnamefont {Y.}~\bibnamefont {Xiong}},\
  }\href {\doibase 10.1103/PhysRevB.76.214204} {\bibfield  {journal} {\bibinfo
  {journal} {Phys. Rev. B}\ }\textbf {\bibinfo {volume} {76}},\ \bibinfo
  {pages} {214204} (\bibinfo {year} {2007})}\BibitemShut {NoStop}%
\bibitem [{\citenamefont {Sheinfux}\ \emph {et~al.}(2017)\citenamefont
  {Sheinfux}, \citenamefont {Lumer}, \citenamefont {Ankonina}, \citenamefont
  {Genack}, \citenamefont {Bartal},\ and\ \citenamefont
  {Segev}}]{Sheinfux2017}%
  \BibitemOpen
  \bibfield  {author} {\bibinfo {author} {\bibfnamefont {H.~H.}\ \bibnamefont
  {Sheinfux}}, \bibinfo {author} {\bibfnamefont {Y.}~\bibnamefont {Lumer}},
  \bibinfo {author} {\bibfnamefont {G.}~\bibnamefont {Ankonina}}, \bibinfo
  {author} {\bibfnamefont {A.~Z.}\ \bibnamefont {Genack}}, \bibinfo {author}
  {\bibfnamefont {G.}~\bibnamefont {Bartal}}, \ and\ \bibinfo {author}
  {\bibfnamefont {M.}~\bibnamefont {Segev}},\ }\href
  {http://science.sciencemag.org/content/356/6341/953?utm_campaign=toc_sci-mag_2017-06-01&et_rid=35068952&et_cid=1359598}
  {\bibfield  {journal} {\bibinfo  {journal} {Science}\ }\textbf {\bibinfo
  {volume} {356}},\ \bibinfo {pages} {953} (\bibinfo {year}
  {2017})}\BibitemShut {NoStop}%
\bibitem [{\citenamefont {Pasek}\ \emph {et~al.}(2017)\citenamefont {Pasek},
  \citenamefont {Orso},\ and\ \citenamefont {Delande}}]{Pasek2016}%
  \BibitemOpen
  \bibfield  {author} {\bibinfo {author} {\bibfnamefont {M.}~\bibnamefont
  {Pasek}}, \bibinfo {author} {\bibfnamefont {G.}~\bibnamefont {Orso}}, \ and\
  \bibinfo {author} {\bibfnamefont {D.}~\bibnamefont {Delande}},\ }\href
  {\doibase 10.1103/PhysRevLett.118.170403} {\bibfield  {journal} {\bibinfo
  {journal} {Physical Review Letters}\ }\textbf {\bibinfo {volume} {118}},\
  \bibinfo {pages} {170403} (\bibinfo {year} {2017})}\BibitemShut {NoStop}%
\bibitem [{\citenamefont {Delande}\ \emph {et~al.}(2017)\citenamefont
  {Delande}, \citenamefont {Morales-Molina},\ and\ \citenamefont
  {Sacha}}]{Delande2017}%
  \BibitemOpen
  \bibfield  {author} {\bibinfo {author} {\bibfnamefont {D.}~\bibnamefont
  {Delande}}, \bibinfo {author} {\bibfnamefont {L.}~\bibnamefont
  {Morales-Molina}}, \ and\ \bibinfo {author} {\bibfnamefont {K.}~\bibnamefont
  {Sacha}},\ }\href {\doibase 10.1103/PhysRevLett.119.230404} {\bibfield
  {journal} {\bibinfo  {journal} {Physical Review Letters}\ }\textbf {\bibinfo
  {volume} {119}},\ \bibinfo {pages} {230404} (\bibinfo {year}
  {2017})}\BibitemShut {NoStop}%
\bibitem [{\citenamefont {{Di Sante}}\ \emph {et~al.}(2017)\citenamefont {{Di
  Sante}}, \citenamefont {Fratini}, \citenamefont {Dobrosavljevi{\'{c}}},\ and\
  \citenamefont {Ciuchi}}]{DiSante2016a}%
  \BibitemOpen
  \bibfield  {author} {\bibinfo {author} {\bibfnamefont {D.}~\bibnamefont {{Di
  Sante}}}, \bibinfo {author} {\bibfnamefont {S.}~\bibnamefont {Fratini}},
  \bibinfo {author} {\bibfnamefont {V.}~\bibnamefont {Dobrosavljevi{\'{c}}}}, \
  and\ \bibinfo {author} {\bibfnamefont {S.}~\bibnamefont {Ciuchi}},\ }\href
  {\doibase 10.1103/PhysRevLett.118.036602} {\bibfield  {journal} {\bibinfo
  {journal} {Physical Review Letters}\ }\textbf {\bibinfo {volume} {118}},\
  \bibinfo {pages} {036602} (\bibinfo {year} {2017})}\BibitemShut {NoStop}%
\bibitem [{\citenamefont {Tikhonov}\ \emph {et~al.}(2016)\citenamefont
  {Tikhonov}, \citenamefont {Mirlin},\ and\ \citenamefont
  {Skvortsov}}]{Tikhonov2016a}%
  \BibitemOpen
  \bibfield  {author} {\bibinfo {author} {\bibfnamefont {K.~S.}\ \bibnamefont
  {Tikhonov}}, \bibinfo {author} {\bibfnamefont {A.~D.}\ \bibnamefont
  {Mirlin}}, \ and\ \bibinfo {author} {\bibfnamefont {M.~A.}\ \bibnamefont
  {Skvortsov}},\ }\href {\doibase 10.1103/PhysRevB.94.220203} {\bibfield
  {journal} {\bibinfo  {journal} {Physical Review B}\ }\textbf {\bibinfo
  {volume} {94}},\ \bibinfo {pages} {220203} (\bibinfo {year} {2016})},\
  \Eprint {http://arxiv.org/abs/1604.05353} {arXiv:1604.05353} \BibitemShut
  {NoStop}%
\bibitem [{\citenamefont {Garc{\'{i}}a-Mata}\ \emph {et~al.}(2017)\citenamefont
  {Garc{\'{i}}a-Mata}, \citenamefont {Giraud}, \citenamefont {Georgeot},
  \citenamefont {Martin}, \citenamefont {Dubertrand},\ and\ \citenamefont
  {Lemari{\'{e}}}}]{Garcia-Mata2016}%
  \BibitemOpen
  \bibfield  {author} {\bibinfo {author} {\bibfnamefont {I.}~\bibnamefont
  {Garc{\'{i}}a-Mata}}, \bibinfo {author} {\bibfnamefont {O.}~\bibnamefont
  {Giraud}}, \bibinfo {author} {\bibfnamefont {B.}~\bibnamefont {Georgeot}},
  \bibinfo {author} {\bibfnamefont {J.}~\bibnamefont {Martin}}, \bibinfo
  {author} {\bibfnamefont {R.}~\bibnamefont {Dubertrand}}, \ and\ \bibinfo
  {author} {\bibfnamefont {G.}~\bibnamefont {Lemari{\'{e}}}},\ }\href {\doibase
  10.1103/PhysRevLett.118.166801} {\bibfield  {journal} {\bibinfo  {journal}
  {Physical Review Letters}\ }\textbf {\bibinfo {volume} {118}},\ \bibinfo
  {pages} {166801} (\bibinfo {year} {2017})}\BibitemShut {NoStop}%
\bibitem [{\citenamefont {Smith}\ \emph {et~al.}(2017)\citenamefont {Smith},
  \citenamefont {Knolle}, \citenamefont {Kovrizhin},\ and\ \citenamefont
  {Moessner}}]{Smith2017}%
  \BibitemOpen
  \bibfield  {author} {\bibinfo {author} {\bibfnamefont {A.}~\bibnamefont
  {Smith}}, \bibinfo {author} {\bibfnamefont {J.}~\bibnamefont {Knolle}},
  \bibinfo {author} {\bibfnamefont {D.}~\bibnamefont {Kovrizhin}}, \ and\
  \bibinfo {author} {\bibfnamefont {R.}~\bibnamefont {Moessner}},\ }\href
  {\doibase 10.1103/PhysRevLett.118.266601} {\bibfield  {journal} {\bibinfo
  {journal} {Physical Review Letters}\ }\textbf {\bibinfo {volume} {118}},\
  \bibinfo {pages} {266601} (\bibinfo {year} {2017})}\BibitemShut {NoStop}%
\bibitem [{\citenamefont {Murphy}\ \emph {et~al.}(2017)\citenamefont {Murphy},
  \citenamefont {Cherkaev},\ and\ \citenamefont {Golden}}]{Murphy2017}%
  \BibitemOpen
  \bibfield  {author} {\bibinfo {author} {\bibfnamefont {N.~B.}\ \bibnamefont
  {Murphy}}, \bibinfo {author} {\bibfnamefont {E.}~\bibnamefont {Cherkaev}}, \
  and\ \bibinfo {author} {\bibfnamefont {K.~M.}\ \bibnamefont {Golden}},\
  }\href {\doibase 10.1103/PhysRevLett.118.036401} {\bibfield  {journal}
  {\bibinfo  {journal} {Physical Review Letters}\ }\textbf {\bibinfo {volume}
  {118}},\ \bibinfo {pages} {036401} (\bibinfo {year} {2017})}\BibitemShut
  {NoStop}%
\bibitem [{\citenamefont {Huckestein}(1995)}]{RevModPhys.67.357}%
  \BibitemOpen
  \bibfield  {author} {\bibinfo {author} {\bibfnamefont {B.}~\bibnamefont
  {Huckestein}},\ }\href {\doibase 10.1103/RevModPhys.67.357} {\bibfield
  {journal} {\bibinfo  {journal} {Rev. Mod. Phys.}\ }\textbf {\bibinfo {volume}
  {67}},\ \bibinfo {pages} {357} (\bibinfo {year} {1995})}\BibitemShut
  {NoStop}%
\bibitem [{\citenamefont {Lee}\ and\ \citenamefont {Fisher}(1981)}]{Lee1981}%
  \BibitemOpen
  \bibfield  {author} {\bibinfo {author} {\bibfnamefont {P.~A.}\ \bibnamefont
  {Lee}}\ and\ \bibinfo {author} {\bibfnamefont {D.~S.}\ \bibnamefont
  {Fisher}},\ }\href {\doibase 10.1103/PhysRevLett.47.882} {\bibfield
  {journal} {\bibinfo  {journal} {Physical Review Letters}\ }\textbf {\bibinfo
  {volume} {47}},\ \bibinfo {pages} {882} (\bibinfo {year} {1981})}\BibitemShut
  {NoStop}%
\bibitem [{\citenamefont {Beenakker}(1997)}]{RevModPhys.69.731}%
  \BibitemOpen
  \bibfield  {author} {\bibinfo {author} {\bibfnamefont {C.~W.~J.}\
  \bibnamefont {Beenakker}},\ }\href {\doibase 10.1103/RevModPhys.69.731}
  {\bibfield  {journal} {\bibinfo  {journal} {Rev. Mod. Phys.}\ }\textbf
  {\bibinfo {volume} {69}},\ \bibinfo {pages} {731} (\bibinfo {year}
  {1997})}\BibitemShut {NoStop}%
\bibitem [{\citenamefont {Brody}\ \emph {et~al.}(1981)\citenamefont {Brody},
  \citenamefont {Flores}, \citenamefont {French}, \citenamefont {Mello},
  \citenamefont {Pandey},\ and\ \citenamefont {Wong}}]{RevModPhys.53.385}%
  \BibitemOpen
  \bibfield  {author} {\bibinfo {author} {\bibfnamefont {T.~A.}\ \bibnamefont
  {Brody}}, \bibinfo {author} {\bibfnamefont {J.}~\bibnamefont {Flores}},
  \bibinfo {author} {\bibfnamefont {J.~B.}\ \bibnamefont {French}}, \bibinfo
  {author} {\bibfnamefont {P.~A.}\ \bibnamefont {Mello}}, \bibinfo {author}
  {\bibfnamefont {A.}~\bibnamefont {Pandey}}, \ and\ \bibinfo {author}
  {\bibfnamefont {S.~S.~M.}\ \bibnamefont {Wong}},\ }\href {\doibase
  10.1103/RevModPhys.53.385} {\bibfield  {journal} {\bibinfo  {journal} {Rev.
  Mod. Phys.}\ }\textbf {\bibinfo {volume} {53}},\ \bibinfo {pages} {385}
  (\bibinfo {year} {1981})}\BibitemShut {NoStop}%
\bibitem [{\citenamefont {{Xiong}}(2021)}]{2021arXiv210400847X}%
  \BibitemOpen
  \bibfield  {author} {\bibinfo {author} {\bibfnamefont {Y.}~\bibnamefont
  {{Xiong}}},\ }\href@noop {} {\bibfield  {journal} {\bibinfo  {journal}
  {arXiv:2104.00847}\ } (\bibinfo {year} {2021})}\BibitemShut {NoStop}%
\bibitem [{\citenamefont {MacKinnon}\ and\ \citenamefont
  {Kramer}(1981)}]{PhysRevLett.47.1546}%
  \BibitemOpen
  \bibfield  {author} {\bibinfo {author} {\bibfnamefont {A.}~\bibnamefont
  {MacKinnon}}\ and\ \bibinfo {author} {\bibfnamefont {B.}~\bibnamefont
  {Kramer}},\ }\href {\doibase 10.1103/PhysRevLett.47.1546} {\bibfield
  {journal} {\bibinfo  {journal} {Phys. Rev. Lett.}\ }\textbf {\bibinfo
  {volume} {47}},\ \bibinfo {pages} {1546} (\bibinfo {year}
  {1981})}\BibitemShut {NoStop}%
\bibitem [{\citenamefont {Pichard}\ and\ \citenamefont
  {Sarma}(1981)}]{Pichard_1981}%
  \BibitemOpen
  \bibfield  {author} {\bibinfo {author} {\bibfnamefont {J.~L.}\ \bibnamefont
  {Pichard}}\ and\ \bibinfo {author} {\bibfnamefont {G.}~\bibnamefont
  {Sarma}},\ }\href {\doibase 10.1088/0022-3719/14/6/003} {\bibfield  {journal}
  {\bibinfo  {journal} {Journal of Physics C: Solid State Physics}\ }\textbf
  {\bibinfo {volume} {14}},\ \bibinfo {pages} {L127} (\bibinfo {year}
  {1981})}\BibitemShut {NoStop}%
\bibitem [{\citenamefont {MacKinnon}\ and\ \citenamefont
  {Kramer}(1983)}]{MacKinnon1983}%
  \BibitemOpen
  \bibfield  {author} {\bibinfo {author} {\bibfnamefont {A.}~\bibnamefont
  {MacKinnon}}\ and\ \bibinfo {author} {\bibfnamefont {B.}~\bibnamefont
  {Kramer}},\ }\href {\doibase 10.1007/BF01578242} {\bibfield  {journal}
  {\bibinfo  {journal} {Zeitschrift f{\"{u}}r Physik B Condensed Matter}\
  }\textbf {\bibinfo {volume} {53}},\ \bibinfo {pages} {1} (\bibinfo {year}
  {1983})}\BibitemShut {NoStop}%
\bibitem [{\citenamefont {Luo}\ \emph {et~al.}(2021)\citenamefont {Luo},
  \citenamefont {Ohtsuki},\ and\ \citenamefont
  {Shindou}}]{PhysRevB.104.104203}%
  \BibitemOpen
  \bibfield  {author} {\bibinfo {author} {\bibfnamefont {X.}~\bibnamefont
  {Luo}}, \bibinfo {author} {\bibfnamefont {T.}~\bibnamefont {Ohtsuki}}, \ and\
  \bibinfo {author} {\bibfnamefont {R.}~\bibnamefont {Shindou}},\ }\href
  {\doibase 10.1103/PhysRevB.104.104203} {\bibfield  {journal} {\bibinfo
  {journal} {Phys. Rev. B}\ }\textbf {\bibinfo {volume} {104}},\ \bibinfo
  {pages} {104203} (\bibinfo {year} {2021})}\BibitemShut {NoStop}%
\bibitem [{\citenamefont {Wigner}(1951)}]{wigner_1951}%
  \BibitemOpen
  \bibfield  {author} {\bibinfo {author} {\bibfnamefont {E.~P.}\ \bibnamefont
  {Wigner}},\ }\href {\doibase 10.1017/S0305004100027237} {\bibfield  {journal}
  {\bibinfo  {journal} {Mathematical Proceedings of the Cambridge Philosophical
  Society}\ }\textbf {\bibinfo {volume} {47}},\ \bibinfo {pages} {790–798}
  (\bibinfo {year} {1951})}\BibitemShut {NoStop}%
\bibitem [{\citenamefont {Dyson}(1962{\natexlab{a}})}]{Dyson}%
  \BibitemOpen
  \bibfield  {author} {\bibinfo {author} {\bibfnamefont {F.~J.}\ \bibnamefont
  {Dyson}},\ }\href {\doibase 10.1063/1.1703773} {\bibfield  {journal}
  {\bibinfo  {journal} {Journal of Mathematical Physics}\ }\textbf {\bibinfo
  {volume} {3}},\ \bibinfo {pages} {140} (\bibinfo {year}
  {1962}{\natexlab{a}})}\BibitemShut {NoStop}%
\bibitem [{\citenamefont {Dyson}(1962{\natexlab{b}})}]{Dyson1}%
  \BibitemOpen
  \bibfield  {author} {\bibinfo {author} {\bibfnamefont {F.~J.}\ \bibnamefont
  {Dyson}},\ }\href {\doibase 10.1063/1.1703863} {\bibfield  {journal}
  {\bibinfo  {journal} {Journal of Mathematical Physics}\ }\textbf {\bibinfo
  {volume} {3}},\ \bibinfo {pages} {1199} (\bibinfo {year}
  {1962}{\natexlab{b}})}\BibitemShut {NoStop}%
\bibitem [{\citenamefont {Santos}\ \emph {et~al.}(2016)\citenamefont {Santos},
  \citenamefont {Borgonovi},\ and\ \citenamefont
  {Celardo}}]{PhysRevLett.116.250402}%
  \BibitemOpen
  \bibfield  {author} {\bibinfo {author} {\bibfnamefont {L.~F.}\ \bibnamefont
  {Santos}}, \bibinfo {author} {\bibfnamefont {F.}~\bibnamefont {Borgonovi}}, \
  and\ \bibinfo {author} {\bibfnamefont {G.~L.}\ \bibnamefont {Celardo}},\
  }\href {\doibase 10.1103/PhysRevLett.116.250402} {\bibfield  {journal}
  {\bibinfo  {journal} {Phys. Rev. Lett.}\ }\textbf {\bibinfo {volume} {116}},\
  \bibinfo {pages} {250402} (\bibinfo {year} {2016})}\BibitemShut {NoStop}%
\bibitem [{\citenamefont {Celardo}\ \emph {et~al.}(2016)\citenamefont
  {Celardo}, \citenamefont {Kaiser},\ and\ \citenamefont
  {Borgonovi}}]{PhysRevB.94.144206}%
  \BibitemOpen
  \bibfield  {author} {\bibinfo {author} {\bibfnamefont {G.~L.}\ \bibnamefont
  {Celardo}}, \bibinfo {author} {\bibfnamefont {R.}~\bibnamefont {Kaiser}}, \
  and\ \bibinfo {author} {\bibfnamefont {F.}~\bibnamefont {Borgonovi}},\ }\href
  {\doibase 10.1103/PhysRevB.94.144206} {\bibfield  {journal} {\bibinfo
  {journal} {Phys. Rev. B}\ }\textbf {\bibinfo {volume} {94}},\ \bibinfo
  {pages} {144206} (\bibinfo {year} {2016})}\BibitemShut {NoStop}%
\bibitem [{\citenamefont {Abanin}\ \emph {et~al.}(2019)\citenamefont {Abanin},
  \citenamefont {Altman}, \citenamefont {Bloch},\ and\ \citenamefont
  {Serbyn}}]{RevModPhys.91.021001}%
  \BibitemOpen
  \bibfield  {author} {\bibinfo {author} {\bibfnamefont {D.~A.}\ \bibnamefont
  {Abanin}}, \bibinfo {author} {\bibfnamefont {E.}~\bibnamefont {Altman}},
  \bibinfo {author} {\bibfnamefont {I.}~\bibnamefont {Bloch}}, \ and\ \bibinfo
  {author} {\bibfnamefont {M.}~\bibnamefont {Serbyn}},\ }\href {\doibase
  10.1103/RevModPhys.91.021001} {\bibfield  {journal} {\bibinfo  {journal}
  {Rev. Mod. Phys.}\ }\textbf {\bibinfo {volume} {91}},\ \bibinfo {pages}
  {021001} (\bibinfo {year} {2019})}\BibitemShut {NoStop}%
\end{thebibliography}%

\end{document}